\documentclass[journal,twoside]{IEEEtran}

\usepackage{cite}
\usepackage{amsmath,amssymb,amsfonts}
\usepackage{graphicx}
\usepackage{float}
\usepackage{booktabs}
\usepackage[caption=false,font=footnotesize]{subfig}
\usepackage{multirow}
\usepackage{array}
\usepackage{siunitx}
\usepackage{algorithm}
\usepackage{algpseudocode}
\usepackage[hidelinks]{hyperref}
\usepackage{xcolor}
\usepackage{orcidlink}
\usepackage[utf8]{inputenc}  % usually default, keep it explicit
\usepackage[T1]{fontenc}
\usepackage{textcomp}        % for \textquote* macros

\DeclareUnicodeCharacter{2011}{\nobreakdash-} % non‑breaking hyphen
\DeclareUnicodeCharacter{202F}{\,}            % narrow no‑break space -> thin space
\DeclareUnicodeCharacter{00A0}{~}             % no‑break space
\DeclareUnicodeCharacter{2013}{--}            % en dash
\DeclareUnicodeCharacter{2014}{---}           % em dash
\DeclareUnicodeCharacter{2026}{\ldots}        % ellipsis
\DeclareUnicodeCharacter{2018}{\textquoteleft}
\DeclareUnicodeCharacter{2019}{\textquoteright}
\DeclareUnicodeCharacter{201C}{\textquotedblleft}
\DeclareUnicodeCharacter{201D}{\textquotedblright}
\DeclareUnicodeCharacter{2212}{-}             % minus sign

\newcommand{\method}{\textsc{TTT-PLC}} % Test-Time Tuning for PLC

\title{Self-Supervised Test-Time Tuning for Packet Loss Concealment}

\author{Yehoshua Dissen\,\orcidlink{0000-0002-9380-8169}, Joseph Keshet\,\orcidlink{0000-0003-2332-5783},~\IEEEmembership{Senior Member,~IEEE}%
\thanks{The authors are with the Andrew and Erna Viterbi Faculty of Electrical and Computer Engineering, Technion--Israel Institute of Technology, Haifa 3200003, Israel (e-mail: yehoshuad@campus.technion.ac.il; jkeshet@technion.ac.il).}
}

\begin{document}
\maketitle

% ---------- Abstract & Index Terms ----------

\begin{abstract}
Packet loss concealment (PLC) reconstructs audio packets that are missing at the receiver, usually with a trained model whose parameters remain fixed at deployment time. This treats the PLC model as static, even though each call or recording exposes signal-specific information through the packets that did arrive. We present TTT-PLC, a self-supervised test-time tuning framework that adapts existing PLC models using only those received packets. The method creates supervision by synthetically masking portions of the available signal, training the model to conceal them with its native PLC objective, and then using the adapted model to reconstruct the true packet losses. No clean reference signal, external adaptation data, or architectural modification is required.

We study TTT-PLC in two deployment settings. In the non-causal setting, the received file is available before reconstruction, allowing repeated self-supervised adaptation passes and providing a per-file adaptation ceiling. In the causal setting, audio is streamed without revising emitted samples; adaptation is performed only on completed past blocks, and updated parameters affect only future audio. We instantiate the framework on two public PLC backbones, FRN, a recurrent full-band speech PLC model, and PARCnet, a hybrid autoregressive-neural model for networked music. Across these settings, the results show that pretrained PLC systems do not need to be treated as fixed at inference time, the still-observed portions of a lossy signal can provide an effective training signal for improving concealment on that same signal.
\end{abstract}
% ====================== 1. Introduction (expanded) ======================

\section{Introduction}
Packet loss concealment (PLC) is the task of reconstructing missing audio packets from the signal that remains available at the receiver. It is a core component of VoIP, conferencing, and networked music systems, where packet-switched transport creates dropouts that may occur as isolated frames or as bursts. Classical concealment modules in codecs such as G.711 and Opus use low-complexity signal extrapolation or decoder-state heuristics \cite{ITU1999G711AppI,Valin2012Opus}. Modern neural PLC models instead learn to predict missing waveform or time-frequency content from context, and recent challenges have provided realistic real-call loss traces and common evaluation protocols \cite{Diener2022PLCChallenge,Diener2023PLCMOS}.

Most learned PLC systems are deployed as frozen predictors: once trained, the same parameters are used for every call, every speaker, and every acoustic condition. This is convenient, but it ignores a useful source of information that exists at inference time. Even a lossy recording contains many received packets. Those packets reveal the file-specific speaker, instrument, room, background, spectral scale, and sometimes the local loss pattern. The central question in this paper is whether a pretrained PLC model can exploit that received signal at test time, without using clean references or any additional training data.

We propose \method, a self-supervised test-time tuning framework for existing PLC backbones. The method constructs an auxiliary PLC task from the received packets themselves: it masks a subset of packets that were actually received, optimizes the pretrained model to reconstruct them using the model's native loss, and then applies the adapted model to the packets that were truly lost. Thus, the supervision comes only from the same corrupted test file on which concealment is evaluated. The downstream task remains PLC; the synthetic masking is only an adaptation mechanism.

We study two deployment settings. In the non-causal setting, the complete received file is available before reconstruction, so the model can be adapted through repeated self-supervised passes over the received signal. This setting provides a per-file adaptation ceiling and is relevant for post-call restoration, archive repair, and buffered processing. In the causal setting, audio is emitted as it arrives and already-emitted samples are never revised. Adaptation is performed only on blocks that have already been received, and updated parameters affect only future blocks. This setting tests whether self-supervised test-time tuning can be made compatible with causal PLC.

We instantiate the framework on two public and complementary PLC backbones. FRN is a recurrent full-band speech PLC model operating at 48 kHz \cite{Nguyen2023FRN}. PARCnet is a real-time hybrid autoregressive-neural PLC model designed for networked music applications \cite{Mezza2024PARCnet}. The two models differ in signal representation, architecture, target domain, and deployment assumptions: FRN targets full-band speech with a recurrent frequency-domain predictor, while PARCnet combines waveform-domain autoregressive prediction with a neural residual model for music. This lets us evaluate whether the same test-time tuning principle applies beyond a single architecture or audio domain.

This paper develops a self-supervised test-time tuning framework for PLC that uses only received packets and does not require new training data, clean references, or changes to the pretrained backbone. We show how the framework can be applied both non-causally, when the received file can be adapted before final reconstruction, and strictly causally, when adaptation is restricted to past audio. For recurrent speech PLC, we introduce a block-replay procedure that performs burst-aware adaptation on completed blocks and applies the updated model only to future audio. For music PLC, we study whether a model trained for one signal class can adapt on the current file when the test signal is in domain or out of domain. We use streaming every-packet adaptation with held-out snapshot selection, so the model is updated from received packets while the deployed checkpoint is chosen without access to the true losses. We evaluate the approach across speech and music backbones, non-causal and causal deployment settings, in-domain and out-of-domain signals, ablations, and runtime measurements. The code and experiment scripts are available publicly. \footnote{\texttt{https://github.com/MLSpeech/TTT-PLC}}

\section{Related Work}

Packet loss concealment (PLC) is the receiver-side task of reconstructing missing audio packets from the packets that arrived. Classical PLC methods were designed for low-complexity, low-latency communication systems. The G.711 Appendix~I standard~\cite{ITU1999G711AppI} uses waveform substitution and attenuation to conceal short erasures. Linear-prediction methods such as Gunduzhan and Momtahan~\cite{Gunduzhan2001LinearPredictionPLC} estimate the missing waveform from the recent past, while waveform extrapolation and overlap-add approaches such as Chen~\cite{Chen2009WaveformPLC} exploit local temporal continuity. Modern codecs also include packet-loss robustness internally; Opus~\cite{Valin2012Opus} combines linear prediction, transform coding, and concealment tools for interactive speech and audio. These signal-processing methods remain attractive because of their speed and stability, but their behavior is fixed by design and cannot specialize to a particular speaker, instrument, recording, or loss pattern at deployment time.

Deep PLC methods learn a data-driven prior for missing audio. Early neural approaches formulated PLC as future-frame prediction or regression from local context. Lin et al.~\cite{Lin2021TimeDomainCRN} use convolutional and recurrent networks for time-domain reconstruction, and Westhausen and Meyer~\cite{Westhausen2022tPLCnet} developed a time-domain PLC model, tPLCnet, that uses recurrent neural networks to predict missing waveform samples from a short causal context. Generative approaches have also been explored: Pascual et al.~\cite{Pascual2021AdversarialPLC} use adversarial auto-encoding for packet-loss recovery, and Wang et al.~\cite{Wang2021TemporalSpectralGANPLC} propose a temporal-spectral GAN to synthesize plausible missing regions. Nguyen et al.~\cite{Nguyen2023FRN} introduced FRN, which performs full-band blind speech PLC with a recurrent frequency-domain model, avoiding the need for an explicit loss mask at inference. For networked music, PARCnet~\cite{Mezza2024PARCnet} combines a classical autoregressive predictor with a neural residual branch to meet low-latency musical constraints. Recent diffusion and flow-based PLC methods, such as Diff-PLC~\cite{Yang2024DiffPLC} and Flow-PLC~\cite{Yang2025FlowPLC}, further improve the expressiveness of neural concealment by modeling richer conditional reconstruction distributions. These methods differ in architecture and training objective, but they are usually used as fixed pretrained systems at test time.

PLC evaluation has also become more standardized. The INTERSPEECH 2022 Audio Deep PLC Challenge~\cite{Diener2022PLCChallenge} introduced a public benchmark with realistic loss traces collected from communication scenarios, and the ICASSP 2024 Audio Deep PLC Challenge~\cite{diener2025icassp} continued this line of evaluation. PLCMOS~\cite{Diener2023PLCMOS} provides a learned perceptual metric targeted specifically at PLC artifacts, complementing traditional intrusive metrics such as PESQ~\cite{Rix2001PESQ} and STOI~\cite{Taal2011STOI}. Controlled experiments often use bursty loss models such as Gilbert--Elliott~\cite{Hasslinger2008GilbertElliott} to approximate clustered packet drops. These benchmarks and metrics are important for test-time adaptation because the amount of useful received signal, the burst structure, and the local loss density can all affect whether adaptation helps or hurts.
Several recent works connect PLC with automatic speech recognition. Yang and Chang~\cite{Yang2023ASRGuided} use an ASR-oriented auxiliary loss to guide PLC toward reconstructions that preserve recognition-relevant information. Zhang et al.~\cite{Zhang2024TDPLC} introduce semantic awareness into PLC, using linguistic context to improve reconstruction during long packet-loss bursts. Dissen et al.~\cite{Dissen2024EnhancedASR,dissen2025front} focused on PLC for improving ASR by training a front-end PLC network guided by a frozen ASR model. 

Self-supervised learning provides a way to train models from structure available in the input rather than from manual labels. Noise2Void~\cite{Krull2019Noise2Void} and Noise2Self~\cite{Batson2019Noise2Self} show that denoisers can be trained by predicting held-out observations from their context.  
Test-time adaptation updates model parameters during inference to better match the current input distribution. Tent~\cite{Wang2021Tent} adapts image classifiers by minimizing prediction entropy on unlabeled test data. Related approaches in speech use unlabeled objectives for zero-shot denoising or dereverberation~\cite{Kim2024ZeroShotTTA}. In our setting, the unlabeled test input has a special structure: some packets are missing and cannot be used for loss computation, while other packets are received and can be used to define a self-supervised concealment task. This differs from standard test-time adaptation, where the input is usually fully observed but unlabeled. It also creates different deployment constraints: in the causal setting, adaptation may use only past received audio and must not revise already-emitted samples.

Deep Image Prior~\cite{Ulyanov2018DeepImagePrior} demonstrates that an untrained convolutional network can regularize inverse problems when optimized to fit the observed pixels of a single corrupted image. Deep Decoder~\cite{Heckel2019DeepDecoder} shows that compact untrained neural architectures can act as image priors. NeRF~\cite{Mildenhall2020NeRF} optimizes a neural radiance field to fit the observations of a single scene, showing the potential of per-instance neural fitting. In inverse problems, ZS-SSL~\cite{yaman2022zero} partitions acquired MRI measurements into disjoint subsets for self-supervised reconstruction without fully sampled reference data. Our approach follows the per-instance optimization philosophy but differs from these methods in two ways: it starts from a pretrained PLC model rather than fitting a new representation from scratch, and it trains on a proxy concealment task constructed from received packets while evaluating on the true packet losses.

\section{Method}
This study proposes a test-time adaptation framework for packet loss concealment that improves pretrained PLC models without changing their architectures or using additional training data. As described earlier, most neural PLC systems are deployed as fixed models: a corrupted audio stream is passed through the pretrained concealer, and the model parameters remain unchanged for every call, recording, or packet-loss pattern. This is limiting because the received portions of the test signal contain information about the speaker, instrument, channel, and local temporal structure of the current file. Here, we use that information directly by constructing a self-supervised PLC task from the packets that were successfully received. Specifically, we synthetically mask a subset of the available packets, train the model to reconstruct them using its native PLC objective, and then use the adapted model to conceal the true packet losses.

We start by presenting the notation and general setting. We use the term packet to denote the unit on which the loss mask is defined. For the FRN model \cite{Nguyen2023FRN} this unit is an STFT frame, whereas for the PARCnet model \cite{Mezza2024PARCnet} this unit is a 10~ms waveform packet. Let $\mathbf{x}_{1:T} = (\mathbf{x}_1,\ldots,\mathbf{x}_T)$ 
denote the clean packet sequence, where \(\mathbf{x}_t\in\mathbb{R}^{N}\), and let \(m_{1:T}\in\{0,1\}^{T}\) denote the packet-loss mask. We use \(m_t=1\) for a true packet loss and \(m_t=0\) for a packet that arrived at the receiver. This defines two disjoint sets. The first set is the set of speech frames which are originally \emph{lost} and is denoted $\mathcal{L}$. The second set is the set of the \emph{observed} frames, denoted by $\mathcal{O}$. That is
\begin{equation}
\begin{aligned}
\mathcal{L} &= \{t:m_t=1\},\\
\mathcal{O} &= \{t:m_t=0\},
\end{aligned}
\end{equation}
where \(\mathcal{L}\cap\mathcal{O}=\emptyset\) and \(\mathcal{L}\cup\mathcal{O}=\{1,\ldots,T\}\). The clean packet \(\mathbf{x}_t\) is available to the receiver only for \(t\in\mathcal{O}\). Frames (or packets) in \(\mathcal{L}\) are used only for final evaluation and are never used as adaptation targets.

Let \(f_\theta\) be a pretrained PLC model with parameters \(\theta\). Given the context available at frame or packet index \(t\), denoted by \(\mathbf{u}_t\), the model predicts the lost packet $\mathbf{x}_t$
\begin{equation}
\hat{\mathbf{x}}_t = f_\theta(\mathbf{u}_t).
\end{equation}
Note that is the packet at time $t$ is observed the model should return it as is. The context \(\mathbf{u}_t\) depends on the backbone. It may contain past waveform samples, STFT frames, recurrent states, an autoregressive buffer, or a combination of these. Let \(\ell(\hat{\mathbf{x}}_t,\mathbf{x}_t)\) denote the native PLC loss used to train and find the parameters $\theta$ of the pretrained model $f_\theta$.

The proposed method creates a self-supervised PLC task from the observed packets. We choose subsets of \(\mathcal{O}\), mask them \emph{synthetically}, and train the model to reconstruct them. The true lost packets \(\mathcal{L}\) remain masked during adaptation and are not used in the loss.

\subsection{Non-causal test-time tuning}

In the non-causal setting, the complete received file is available before final reconstruction. We adapt a separate copy of the pretrained model for each test file and reset the model to the public checkpoint before moving to the next file.

We first partition the observed packets into a held-out validation set and an adaptation pool. Specifically, we sample a \emph{held-out} set $\mathcal{H}\subset\mathcal{O}$, where \(\mathcal{H}\) contains packets that arrived at the receiver but are not used for gradient updates. The remaining observed packets form the adaptation pool, $\mathcal{A}=\mathcal{O}\setminus\mathcal{H}$. Thus, in the non-causal setting, the three sets \(\mathcal{L}\), \(\mathcal{H}\), and \(\mathcal{A}\) are disjoint, and $\mathcal{L}\cup\mathcal{H}\cup\mathcal{A}=\{1,\ldots,T\}$.

During adaptation, the pool \(\mathcal{A}\) is further split into synthetic masking folds. For fold \(k\), we choose a synthetic target set \(\mathcal{C}_k\subset\mathcal{A}\), and define the corresponding context set
$\mathcal{D}_k=\mathcal{A}\setminus\mathcal{C}_k$.
For every fold, the sets \(\mathcal{L}\), \(\mathcal{H}\), \(\mathcal{C}_k\), and \(\mathcal{D}_k\) are pairwise disjoint, and their union is the complete packet index set. The packets in \(\mathcal{C}_k\) are \emph{synthetically hidden and used as self-supervised training targets}. The packets in \(\mathcal{D}_k\) remain available as context. The held-out packets in \(\mathcal{H}\) are also masked during training, but they are not used in the training loss.

For any packet index set \(\mathcal{Z}\subseteq\{1,\ldots,T\}\), we define the \emph{masking} operator
\begin{equation}
\mathcal{M}_{\mathcal{Z}}(\mathbf{x}_t) =
\begin{cases}
\mathbf{0}, & t\in\mathcal{Z},\\
\mathbf{x}_t, & t\notin\mathcal{Z}.
\end{cases}
\end{equation}
The training objective for fold \(k\) is
\begin{equation}
\mathcal{L}_{\mathrm{train}}^{(k)}(\theta)
=
\sum_{t\in\mathcal{C}_k}
\ell\!\left(
f_\theta\!\left(\mathbf{u}_t^{(k)}\right),
\mathbf{x}_t
\right),
\end{equation}
where the context \(\mathbf{u}_t^{(k)}\) is computed from the synthetically corrupted waveform $\mathcal{M}_{\mathcal{L}\cup\mathcal{H}\cup\mathcal{C}_k}(\mathbf{x})$. The model therefore sees the original true losses, the held-out validation packets, and the current synthetic training targets as missing. The loss is evaluated only on \(\mathcal{C}_k\), which contains packets that were actually received.

For early stopping or model selection, we evaluate the held-out objective
\begin{equation}
\mathcal{L}_{\mathrm{val}}(\theta)
=
\sum_{t\in\mathcal{H}}
\ell\!\left(
f_\theta\!\left(\mathbf{v}_t\right),
\mathbf{x}_t
\right),
\end{equation}
where the context \(\mathbf{v}_t\) is computed from $\mathcal{M}_{\mathcal{L}\cup\mathcal{H}}(\mathbf{x})$. Validation therefore uses the same principle as training: packets in \(\mathcal{H}\) are treated as synthetic losses, and their known received values provide the validation targets. The final adapted checkpoint \(\theta^\star\) is selected using \(\mathcal{L}_{\mathrm{val}}(\theta)\), not the true missing packets.

After adaptation, the model reconstructs the true packet losses using the received waveform with only the true losses masked:
\begin{equation}
\hat{\mathbf{x}}_t =
f_{\theta^\star}(\mathbf{w}_t),
\qquad
 t\in\mathcal{L},
\end{equation}
where the context \(\mathbf{w}_t\) is computed from $\mathcal{M}_{\mathcal{L}}(\mathbf{x})$. Algorithm~\ref{alg:noncausal} summarizes the non-causal procedure.

\begin{algorithm}[t]
\caption{\method\ non-causal test-time tuning}
\label{alg:noncausal}
\begin{algorithmic}[1]
\Require Test packets \(\mathbf{x}_{1:T}\), loss mask \(m_{1:T}\), pretrained weights \(\theta_0\)
\State Define true losses \(\mathcal{L}=\{t:m_t=1\}\) and observed packets \(\mathcal{O}=\{t:m_t=0\}\)
\State Sample held-out set \(\mathcal{H}\subset\mathcal{O}\) and set \(\mathcal{A}=\mathcal{O}\setminus\mathcal{H}\)
\State Initialize \(\theta\leftarrow\theta_0\)
\For{epoch \(=1,\ldots,E\)}
  \State Form synthetic masking folds over \(\mathcal{A}\)
  \For{each fold \(k\)}
    \State Choose target set \(\mathcal{C}_k\subset\mathcal{A}\) and context set \(\mathcal{D}_k=\mathcal{A}\setminus\mathcal{C}_k\)
    \State Mask \(\mathcal{L}\cup\mathcal{H}\cup\mathcal{C}_k\)
    \State Update \(\theta\) using targets in \(\mathcal{C}_k\)
  \EndFor
  \State Compute \(\mathcal{L}_{\mathrm{val}}(\theta)\) on \(\mathcal{H}\) and keep the best checkpoint \(\theta^\star\)
\EndFor
\State Reconstruct the true losses \(t\in\mathcal{L}\) using \(\theta^\star\)
\end{algorithmic}
\end{algorithm}

\subsection{Causal test-time tuning}

In the causal setting, audio is emitted as it arrives and already-emitted samples are never revised. The adaptation process must therefore satisfy two constraints. First, the output at time \(t\) may depend only on packets received before or at \(t\). Second, any parameter update performed after time \(t\) may affect only future outputs.

A causal update can be applied whenever an observed packet arrives:
\begin{equation}
\theta_{t+1}
=
\theta_t
-
\eta\nabla_\theta
\ell\!\left(
f_{\theta_t}(\mathbf{u}_t),
\mathbf{x}_t
\right),
\qquad t\in\mathcal{O}.
\end{equation}
No loss is applied for \(t\in\mathcal{L}\), since the clean packet is unavailable. However, single-packet updates are not always sufficient for PLC. In bursty packet loss, the model may advance through several consecutive missing packets while feeding its own predictions back into the future context. A synthetic single-packet mask does not reproduce this inference condition.

For recurrent PLC models such as FRN, we therefore use burst-aware replay. Instead of masking only one observed packet, a training step masks a short run of observed packets, rolls the model through the entire synthetic burst, and backpropagates through the multi-step prediction trajectory. This makes the self-supervised training condition closer to the actual burst-concealment condition encountered at inference.

To preserve causality, burst-aware training is performed only on completed blocks. Let \(S\) denote the block size. The receiver first streams a block of \(S\) packets using the current model parameters. Received packets are passed through, lost packets are concealed, and the emitted samples are fixed. After the block has ended, the model performs one or more replay epochs using only the packets and loss mask from that completed block. The updated parameters are then used for the next block. Because training on a block occurs only after that block has already been emitted, replay cannot change past outputs and cannot use future packets. Algorithm~\ref{alg:causalreplay} gives the causal block-replay procedure.

\begin{algorithm}[t]
\caption{Causal block replay}
\label{alg:causalreplay}
\begin{algorithmic}[1]
\Require Streamed packets \(\mathbf{x}_{1:T}\), loss mask \(m_{1:T}\), pretrained weights \(\theta_0\), block size \(S\)
\State Initialize live weights \(\theta\leftarrow\theta_0\) and EMA weights \(\bar{\theta}\leftarrow\theta_0\)
\For{each block}
  \State Stream the next \(S\) packets with weights \(\bar{\theta}\)
  \State Pass through received packets and conceal lost packets
  \State Store the completed block and its received/lost mask
  \State Form self-supervised targets by synthetically masking received packets inside the completed block
  \For{replay epoch \(=1,\ldots,E\)}
    \State Roll through burst-masked received packets and update \(\theta\) using the native PLC loss
    \State Update \(\bar{\theta}\leftarrow\rho\bar{\theta}+(1-\rho)\theta\)
  \EndFor
  \State Use \(\bar{\theta}\) only for future blocks; already emitted samples are unchanged
\EndFor
\end{algorithmic}
\end{algorithm}

\subsection{Streaming PARCnet test-time tuning}
\label{subsec:parcnet_streaming_method}

PARCnet requires one additional detail beyond the general masking construction above because its native prediction target extends beyond the packet being concealed. Let \(D=320\) be the packet length and let \(R=80\) be the prediction tail used for cross-fading. For packet \(t\), PARCnet predicts a window of
\begin{equation}
P=D+R=400 .
\end{equation}
We denote the corresponding target window by \(\mathbf{z}_t\in\mathbb{R}^{P}\), the model prediction by \(\hat{\mathbf{z}}_t=f_\theta(\mathbf{u}_t)\), and the packet index containing sample \(j\) of this window by \(p(t,j)\).

For PARCnet, we use a streaming every-packet adaptation procedure. Before adaptation, the observed packets are split into the training set \(\mathcal{A}=\mathcal{O}\setminus\mathcal{H}\) and the held-out validation set \(\mathcal{H}\). The stream buffer is initialized with the lossy signal: received packets contain their observed samples, and true lost packets are zero-filled until they are concealed. The model then scans the waveform once from left to right. If \(t\in\mathcal{L}\), the packet is a true loss; PARCnet conceals it, writes the prediction into the running buffer, and takes no gradient step. If \(t\in\mathcal{A}\), the packet was received and is used for self-supervised adaptation: we mask it from the neural input, predict it using the current running buffer, and update the model using the native PARCnet loss. If \(t\in\mathcal{H}\), the packet remains available in the stream buffer as received audio, but it is never used as a training target.

The PARCnet loss is the released training objective,
\begin{equation}
\ell_{\mathrm{PARC}}
=
100\,\ell_{\mathrm{MSE}}^{\mathbf{q}}
+
0.5\,\ell_{\mathrm{MR\text{-}STFT}}^{\mathbf{q}},
\end{equation}
where both terms are computed with a sample-level loss mask \(\mathbf{q}\), and the squared-error term is normalized per packet. The sample mask is essential because the \(400\)-sample prediction window includes an \(80\)-sample tail that can overlap the next packet. If that next packet is a true loss, those samples are not available to the decoder and must not contribute to the adaptation loss. For a training packet \(t\in\mathcal{A}\), we therefore use
\begin{equation}
q_{t,j}^{\mathrm{tr}}
=
\mathbb{I}\{p(t,j)\in\mathcal{A}\},
\qquad
j=1,\ldots,P .
\end{equation}
Thus, only samples that belong to packets used for training contribute to the loss. Samples from true losses \(\mathcal{L}\) and held-out validation packets \(\mathcal{H}\) have zero weight in the training objective, even if they appear inside the prediction tail.

Every fixed number of updates, we evaluate the current weights on the held-out packets. For validation, packets in \(\mathcal{H}\) are synthetically concealed and scored against their known received samples using the packet NMSE used in the PARCnet evaluation protocol. The validation mask is
\begin{equation}
q_{t,j}^{\mathrm{val}}
=
\mathbb{I}\{p(t,j)\in\mathcal{H}\},
\qquad
t\in\mathcal{H},
\quad
j=1,\ldots,P ,
\end{equation}
and the validation score for a snapshot \(\theta_s\) is
\begin{equation}
V(\theta_s)
=
\frac{1}{|\mathcal{H}|}
\sum_{t\in\mathcal{H}}
10\log_{10}
\frac{
\left\|\mathbf{q}_{t}^{\mathrm{val}}\odot
\left(\hat{\mathbf{z}}_{t}-\mathbf{z}_{t}\right)\right\|_2^2
}{
\left\|\mathbf{q}_{t}^{\mathrm{val}}\odot\mathbf{z}_{t}\right\|_2^2
} .
\end{equation}
The deployed checkpoint is selected by
\begin{equation}
\theta^\star=\theta_{s^\star},
\qquad
s^\star=\operatorname*{arg\,min}_{s} V(\theta_s) .
\end{equation}
The true lost packets \(\mathcal{L}\) are excluded from both training and validation; they are used only for final evaluation. Algorithm~\ref{alg:parcnetstream} summarizes the procedure.

\begin{algorithm}[t]
\caption{PARCnet streaming adaptation with held-out snapshot selection}
\label{alg:parcnetstream}
\begin{algorithmic}[1]
\Require Lossy packets, loss mask \(m_{1:T}\), pretrained weights \(\theta_0\), held-out set \(\mathcal{H}\), snapshot cadence \(C\)
\State Define \(\mathcal{L}=\{t:m_t=1\}\), \(\mathcal{O}=\{t:m_t=0\}\), and \(\mathcal{A}=\mathcal{O}\setminus\mathcal{H}\)
\State Initialize \(\theta\leftarrow\theta_0\), initialize the running buffer with received packets and zeros at true losses
\State Set \(V^\star\leftarrow\infty\) and \(\theta^\star\leftarrow\theta_0\)
\For{\(t=1,\ldots,T\)}
  \If{\(t\in\mathcal{L}\)}
    \State Conceal packet \(t\), write the prediction into the running buffer, and take no gradient step
  \ElsIf{\(t\in\mathcal{A}\)}
    \State Synthetically mask packet \(t\) from the neural input
    \State Predict \(\hat{\mathbf{z}}_t=f_\theta(\mathbf{u}_t)\) using the current running buffer
    \State Update \(\theta\) using \(\ell_{\mathrm{PARC}}\) with training mask \(\mathbf{q}_{t}^{\mathrm{tr}}\)
  \Else
    \State Pass through held-out packet \(t\) and take no gradient step
  \EndIf
  \If{\(C\) gradient updates have occurred since the last snapshot}
    \State Compute \(V(\theta)\) by synthetically concealing \(\mathcal{H}\)
    \If{\(V(\theta)<V^\star\)}
      \State Set \(V^\star\leftarrow V(\theta)\) and \(\theta^\star\leftarrow\theta\)
    \EndIf
  \EndIf
\EndFor
\State For two-pass deployment, reconstruct the true losses \(\mathcal{L}\) using the selected checkpoint \(\theta^\star\)
\end{algorithmic}
\end{algorithm}

\section{Evaluation Settings}
\label{sec:evaluation}

This section describes the pretrained backbones, datasets, packet-loss conditions, metrics, and implementation details used in the experiments. In all experiments, adaptation is performed independently for each test waveform. The model is reset to the released public checkpoint before processing the next waveform, so no information is transferred across evaluation utterances or recordings.

\subsection{Backbones}
\label{subsec:backbones}

We evaluate \method\ on two open-source PLC backbones. The first is FRN~\cite{Nguyen2023FRN}, a full-band recurrent speech PLC model that operates on STFT frames at 48~kHz. FRN processes the signal frame by frame while maintaining recurrent state, and available frames are passed through in the STFT domain. Only frames marked as lost are replaced by the model prediction. For FRN, we continue training with the released spectral objective and evaluate both the non-causal adaptation procedure in Algorithm~\ref{alg:noncausal} and the causal block-replay procedure in Algorithm~\ref{alg:causalreplay}.

The second backbone is PARCnet~\cite{Mezza2024PARCnet}, a hybrid autoregressive-neural PLC model designed for real-time networked music. We use the released PARCnet checkpoint and retain the original AR-plus-neural architecture. Audio is mono at 32~kHz, each packet contains \(D=320\) samples, corresponding to 10~ms, and the model predicts a \(D+80=400\)-sample window so that the final 80 samples can be cross-faded against the subsequent buffer. The frozen PARCnet system combines a 10-packet AR context with a 7-packet neural residual context. The neural residual is faded in over 16 samples, and the combined AR-plus-neural prediction is faded out over the 80-sample tail. Available packets are passed through unchanged, and only packets indicated as lost by the trace are reconstructed.

\subsection{Datasets}
\label{subsec:datasets}

We use separate speech and music evaluation settings to test long-file adaptation, short-file speech PLC, in-domain music behavior, and out-of-domain adaptation.

\emph{LibriSpeech-40.} LibriSpeech \texttt{test-clean}~\cite{Panayotov2015LibriSpeech} is used to test whether adaptation can accumulate over long speech recordings. We concatenate utterances by speaker to form 40 approximately 8-minute files at 48~kHz. This setting is intentionally long: it exposes whether causal adaptation improves after several received blocks and whether the adapted model drifts later in the file. The same 40-speaker set is also used as the out-of-domain speech condition for PARCnet.

\emph{PLC Challenge 2022 speech.} The PLC Challenge 2022 test set contains 966 short blind-test files at 16~kHz with real packet-loss traces from Microsoft Teams calls~\cite{Diener2022PLCChallenge}. We use this benchmark because it is a standard speech PLC evaluation set and because the files are short, which stresses the low-latency setting where a causal replay method has limited time to adapt.

\emph{PARCnet in-domain music.} The in-domain music condition uses the MAESTRO recording used in the PARCnet paper~\cite{hawthorne2018enabling,Mezza2024PARCnet}. We evaluate the full 50-minute file, a 5-minute crop, and five 30-second chunks at 32~kHz. PARCnet was trained on music data and is already well matched to this audio. This condition tests whether test-time tuning can avoid harmful over-adaptation when the frozen model is already close to the target signal distribution.

\emph{PARCnet out-of-domain speech.} To evaluate adaptation under distribution shift, we apply PARCnet to LibriSpeech-40 using the same 32~kHz packet-loss interface. This setting is deliberately out of domain for PARCnet and tests whether self-supervised adaptation from received packets can move a music-trained PLC model toward a new signal class without external data.

\subsection{Packet-loss conditions}
\label{subsec:loss_conditions}

For the LibriSpeech-40, FRN experiments, loss masks are tiled from PLC Challenge 2022 traces. The selected masks span loss rates from 0\% to 70.5\% and contain bursts from 1 to 28 frames. For the PLC Challenge 2022 evaluation, we use the official packet-loss traces as provided with the benchmark. These traces are used without modification.

For the PARCnet experiments, we use nominal packet-loss rates of 10\%, 20\%, and 30\%. Traces are generated once for each file and loss rate using a fixed seed. The true lost packets are known as loss locations, but their clean samples are excluded from training and validation and are used only for final evaluation. The PARCnet evaluation grid contains 141 file-rate cells: 40 LibriSpeech speakers times three loss rates, the 50-minute MAESTRO file times three loss rates, the 5-minute MAESTRO crop times three loss rates, and five 30-second MAESTRO chunks times three loss rates.

\subsection{Metrics and aggregation}
\label{subsec:metrics}

For FRN, we follow the evaluation protocol of the original FRN study and report STOI~\cite{Taal2011STOI}, wideband PESQ~\cite{Rix2001PESQ}, log-spectral distance (LSD), intrusive PLCMOS, and non-intrusive PLCMOS~\cite{Diener2023PLCMOS}. PESQ and PLCMOS are evaluated at 16~kHz. STOI and LSD are computed at the native sampling rate of each corpus. On LibriSpeech-40, all five metrics are computed on non-overlapping 20-second windows. The window scores are first averaged within each file and then averaged across the 40 speakers. The 20-second windowing avoids PLCMOS length limits and keeps the metric protocol consistent across all five metrics. On PLC Challenge 2022, the files are short enough to evaluate as complete files.

For PARCnet, we follow the objective evaluation protocol used in the original PARCnet study. The primary reconstruction metric is packet NMSE in dB, computed over true lost packets as
\begin{equation}
\mathrm{NMSE}_{\mathrm{pkt}}
=
10\log_{10}
\frac{\lVert \mathbf{y}-\hat{\mathbf{y}} \rVert_2^2}
{\lVert \mathbf{y} \rVert_2^2},
\end{equation}
where \(\mathbf{y}\) is the ground-truth lost packet and \(\hat{\mathbf{y}}\) is the reconstructed packet. Lower NMSE is better. We also report Mel spectral convergence (Mel-SC)~\cite{kubichek1993mel}, which measures spectral error after applying a Mel-frequency filterbank to the magnitude STFT. Following PARCnet, Mel-SC is computed using 64 triangular Mel filters and a 512-sample Hann window with 50\% overlap at the native 32~kHz sampling rate. To include perceptual measures, we report PEAQ~\cite{thiede2000peaq} and PLCMOS~\cite{Diener2023PLCMOS}. PEAQ is reported as Objective Difference Grade (ODG), averaged over non-overlapping 10-second windows after resampling to 48~kHz. PLCMOS is the non-intrusive PLCMOS score, averaged over non-overlapping 10-second windows after resampling to 16~kHz. We also track the silence-energy ratio as a diagnostic guardrail for silence-fill artifacts. When confidence intervals are reported, they are 95\% bootstrap intervals over per-file deltas.

\subsection{Implementation details}
\label{subsec:implementation_details}

For FRN, the non-causal setting uses the full received file before final reconstruction and provides a per-file adaptation ceiling. The causal setting uses block replay. In this setting, the model streams a block with the current parameters, emits the corresponding audio, and then trains only on that completed block. Updated weights affect only future blocks.

For LibriSpeech-40, causal FRN uses blocks of 3000 frames, corresponding to 30-seconds, after a 1000-frame warmup. For PLC Challenge 2022, causal FRN uses blocks of 200 frames after a 200-frame warmup, matching the shorter file duration. In both datasets, replay uses six epochs per block, a 128-frame loss history, AdamW with learning rate \(5\times10^{-6}\), weight decay \(10^{-6}\), gradient clipping at L2 norm 2.0, and an L2 pull of strength \(5\times10^{-7}\) on the joiner parameters. The burst depth is anchored to observed packet losses,
\begin{equation}
K_{\mathrm{burst}}=\min\left(12,\max\left(4,K_{\mathrm{obs}}\right)\right),
\end{equation}
where \(K_{\mathrm{obs}}\) is the longest of the last ten completed real bursts. We maintain exponential moving average (EMA) weights with coefficient 0.99 and use this smoothed copy of the adapted model for streaming output.

For PARCnet, the proposed method is streaming every-packet TTT with held-out snapshot selection. We use AdamW with learning rate \(5\times10^{-6}\), weight decay \(10^{-4}\), one update per available training packet, gradient clipping at L2 norm 2.0, and the released PARCnet loss \(100\cdot\mathrm{MSE}+0.5\cdot\mathrm{MR\text{-}STFT}\) with per-packet normalization. A random 10\% subset of available packets is held out before training using seed 42. These held-out packets are never used for gradient updates; they are used only to select checkpoints using held-out packet NMSE. We evaluate snapshots every \(C\) updates, with \(C=400\) for the 40-speaker headline experiments and \(C=3200\) for the 50-minute MAESTRO file. The same adaptation trajectory supports a causal best-val mode, which uses only snapshots validated so far, and a two-pass best-val mode, which selects the best snapshot over the whole file and reconstructs the true losses in a second pass. We also report single-pass oracle snapshot selection as an upper bound and multi-epoch adaptation as a quality-oriented configuration.

The PARCnet prediction window contains the target packet and an 80-sample tail. Because this tail can overlap a neighboring true loss, all PARCnet adaptation losses are computed with a sample-level mask that excludes any samples belonging to true lost packets. This mask is applied to both the MSE and MR-STFT terms. Thus, clean audio at true loss positions is never used during training or validation, even when it lies inside the prediction tail of an observed packet.

% Results
% 

\section{Results}
\label{sec:results}

\begin{table*}[t]
\centering
\caption{FRN results on LibriSpeech-40 long files. Entries are absolute scores; parentheses give the change from frozen and the number of improved files. The causal confidence row gives 95\% bootstrap confidence intervals for the change from frozen.}
\label{tab:frn_libri_unified}
\resizebox{\textwidth}{!}{%
\begin{tabular}{lccccc}
\toprule
Method & STOI $\uparrow$ & PESQ $\uparrow$ & LSD $\downarrow$ & PLCMOS-I $\uparrow$ & PLCMOS-NI $\uparrow$ \\
\midrule
Frozen FRN & 0.911 & 2.442 & 1.036 & 2.661 & 2.589 \\
Non-causal & 0.920 (+0.0092; 39/40) & 2.489 (+0.047; 38/40) & 0.737 (-0.299; 39/40) & 2.706 (+0.045; 32/40) & 2.631 (+0.042; 31/40) \\
Causal replay & 0.919 (+0.0078; 39/40) & 2.482 (+0.040; 38/40) & 0.732 (-0.304; 39/40) & 2.696 (+0.035; 32/40) & 2.626 (+0.037; 30/40) \\
Causal 95\% CI & [+0.0054,+0.0104] & [+0.0313,+0.0485] & [-0.4047,-0.2153] & [+0.0164,+0.0524] & [+0.0228,+0.0518] \\
Causal/non-causal gain & 85\% & 84\% & 102\% & 77\% & 88\% \\
\bottomrule
\end{tabular}%
}
\end{table*}

In this section, we evaluate our method, \method, across speech and music PLC, non-causal and causal deployment settings, in-domain and out-of-domain signals, and multiple packet-loss rates. We first study FRN on long LibriSpeech recordings and on the PLC Challenge 2022 test set to measure how much of the non-causal adaptation gain can be recovered under causality. We then evaluate PARCnet on out-of-domain speech and in-domain music.

\subsection{FRN: non-causal ceiling and causal replay on long files}
\label{subsec:frn_long_results}

Table~\ref{tab:frn_libri_unified} reports FRN results on LibriSpeech-40. The non-causal setting adapts using the full received file before the final reconstruction, and therefore serves as a per-file adaptation ceiling. Causal replay is more constrained: it emits each block before replay training is performed, and updated parameters can affect only future blocks. Because FRN is recurrent and real losses often occur in bursts, replay training uses short synthetic bursts rather than only isolated single-frame masks; we refer to this part of the adaptation as burst-aware. Despite the causal restriction, causal replay recovers most of the non-causal gain. It reaches 84--85\% of the non-causal improvement on STOI and PESQ, 77--88\% on the two PLCMOS heads, and slightly exceeds the non-causal LSD gain. The bootstrap confidence intervals in Table~\ref{tab:frn_libri_unified} show that these gains are consistent across files.

\begin{figure}[t]
\centering
\includegraphics[width=\columnwidth]{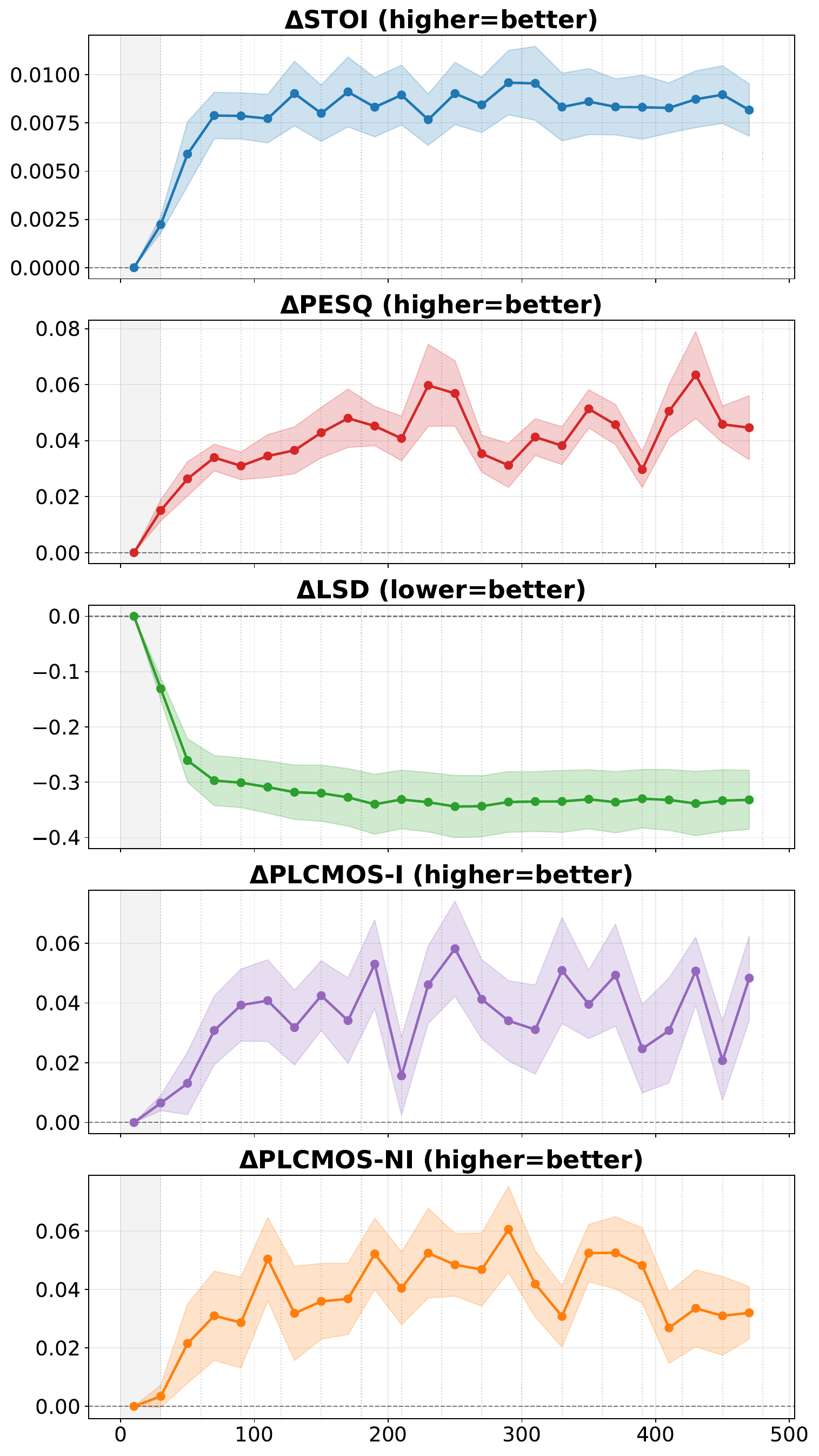}
\caption{Causal FRN block replay improves after the first completed block and remains stable over time. Curves show per-window metric deltas relative to frozen FRN on LibriSpeech-40 speaker files; shaded bands denote \(\pm\) one standard error. The first 30~s block is shaded because no replay has fired yet, so the output matches the frozen baseline. Dotted vertical lines mark subsequent 30~s block boundaries; each completed block is replayed for six self-supervised epochs before updating the weights for future blocks.}
\label{fig:frn_causal_over_time}
\end{figure}

Figure~\ref{fig:frn_causal_over_time} shows how the causal gains evolve over time in the long LibriSpeech-40 files. During the first 30-seconds block, the output is identical to frozen FRN because no replay update has occurred. After the first completed block is replayed, the next block immediately improves across the metrics. By roughly 60--70 seconds, the method reaches a stable plateau and later block boundaries provide smaller refinements, with no visible late-file drift.

\subsection{FRN: PLC Challenge 2022 test set}
\label{subsec:frn_plc22_results}

Table~\ref{tab:frn_plc22_unified} evaluates the same FRN adaptation idea on the 966-file PLC Challenge 2022 test set. This benchmark contains much shorter files than LibriSpeech-40, so the amount of causal adaptation time is limited. Even in this setting, causal replay improves every metric relative to frozen FRN.

\begin{table*}[t]
\centering
\caption{FRN results on the 966-file PLC Challenge 2022 test set. Entries are absolute scores; parentheses give the change from frozen and the win rate. The causal confidence row gives 95\% bootstrap confidence intervals for the change from frozen.}
\label{tab:frn_plc22_unified}
\resizebox{\textwidth}{!}{%
\begin{tabular}{lccccc}
\toprule
Method & STOI $\uparrow$ & PESQ $\uparrow$ & LSD $\downarrow$ & PLCMOS-I $\uparrow$ & PLCMOS-NI $\uparrow$ \\
\midrule
Frozen FRN & 0.8854 & 2.2816 & 2.1756 & 2.7732 & 2.6672 \\
Causal replay & 0.8910 (+0.0056; 70\%) & 2.3706 (+0.089; 86\%) & 1.9776 (-0.198; 79\%) & 2.9092 (+0.136; 74\%) & 2.9352 (+0.268; 81\%) \\
Causal 95\% CI & [+0.0040,+0.0078] & [+0.0786,+0.1001] & [-0.2178,-0.1782] & [+0.1199,+0.1529] & [+0.2390,+0.2966] \\
\bottomrule
\end{tabular}%
}
\end{table*}

The confidence intervals in Table~\ref{tab:frn_plc22_unified} show that the improvements are stable across the 966-file test set. Together with the LibriSpeech-40 results, this shows that burst-aware block replay on received packets improves FRN in both long-file and short-file speech PLC settings.

\subsection{FRN runtime}
\label{subsec:frn_runtime_results}

Table~\ref{tab:frn_runtime} reports FRN timing on a representative LibriSpeech-style file. Here, EMA denotes the exponential moving average weight copy used for reconstruction. The absolute real-time factors come from an unoptimized frame-by-frame Python implementation and should be interpreted mainly as relative costs. Causal replay has a wall-clock cost comparable to the six-epoch non-causal configuration because both perform repeated replay training. Its advantage in this implementation is causal validity rather than lower compute: emitted samples are never revised, and each replay update affects only future blocks.

\begin{table}[t]
\centering
\caption{FRN runtime on a LibriSpeech-style file, single GPU. RTF is compute time divided by audio duration; lower is faster.}
\label{tab:frn_runtime}
\begin{tabular}{lcc}
\toprule
Configuration & RTF $\downarrow$ & Relative RT \\
\midrule
Frozen FRN & 0.31 & 3.2$\times$ \\
Single-pass causal & 0.66 & 1.5$\times$ \\
Non-causal anchored, no EMA & 4.34 & 0.23$\times$ \\
Non-causal anchored+EMA & 4.70 & 0.21$\times$ \\
% \(K=4\), 6 epochs & 3.77 & 0.27$\times$ \\
\bottomrule
\end{tabular}
\end{table}

\subsection{PARCnet: per-file test-time training}
\label{subsec:parcnet_main_results}

We next evaluate \method\ on PARCnet. The adaptation procedure follows the no-leak construction in Section~\ref{subsec:parcnet_streaming_method}: only received packets are used for training and held-out validation, while true lost packets are used only for final evaluation. We use LibriSpeech-40 as an out-of-domain speech condition and the MAESTRO solo-piano recording from the PARCnet evaluation as the in-domain music condition.

We compare frozen PARCnet with three held-out-selection variants. \emph{Causal best-val} conceals each lost packet using the best snapshot validated so far in the stream. \emph{Two-pass best-val} selects the best snapshot after a full adaptation pass and then reconstructs the file with that checkpoint. \emph{Multi-epoch best-val} extends the same selection rule across multiple adaptation epochs and represents the offline, quality-oriented setting. We also report an oracle selector that chooses the best snapshot using the true lost packets; this is an upper bound and is not deployable.

\subsubsection{Out-of-domain and in-domain evaluation}
\label{sec:parcnet-results}

Table~\ref{tab:ood} reports the out-of-domain LibriSpeech results. Frozen PARCnet gives little packet-NMSE improvement relative to zero-fill, ranging from \(-0.06\) to \(+0.30\)~dB across the three packet-loss rates. Causal best-val and two-pass best-val both substantially reduce packet NMSE. Two-pass best-val reaches \(-1.84\), \(-1.76\), and \(-1.59\)~dB at 10\%, 20\%, and 30\% packet loss, respectively, while causal best-val reaches \(-1.63\), \(-1.57\), and \(-1.39\)~dB. Multi-epoch best-val gives the largest packet-NMSE reductions in this condition, reaching \(-2.46\), \(-2.34\), and \(-2.04\)~dB.

Table~\ref{tab:ind} reports the in-domain MAESTRO results on the 50-minute piano file. Frozen PARCnet is already strong in this condition, with packet NMSE between \(-4.28\) and \(-5.28\)~dB across packet-loss rates. The single-pass held-out selectors remain close to frozen and provide modest additional packet-NMSE improvements: causal best-val reaches \(-5.52\), \(-5.12\), and \(-4.40\)~dB at 10\%, 20\%, and 30\% loss, respectively, while two-pass best-val reaches \(-5.59\), \(-5.08\), and \(-4.43\)~dB. The single-pass oracle is only slightly better, indicating that held-out selection leaves little single-pass selection gain unused in this in-domain setting. Multi-epoch adaptation further reduces packet NMSE at all three loss rates, reaching \(-5.94\), \(-5.49\), and \(-4.55\)~dB, although the perceptual metrics remain mixed. Thus, the MAESTRO results show that held-out adaptation can improve packet-level reconstruction even when the frozen in-domain model is already strong, with the largest gains obtained by the offline multi-epoch setting.

\subsubsection{Held-out snapshot selection}
\label{sec:parcnet-oracle}

To evaluate the quality of held-out model selection, we compare best-val selection with an oracle selector that chooses, from the same single-pass adaptation trajectory, the snapshot minimizing packet NMSE on the true lost packets. The oracle uses clean loss-region audio and is therefore not deployable; it is used only as an upper bound on snapshot selection within a fixed adaptation trajectory. Across all 141 file-rate cells, held-out selection is within a mean 0.02~dB of this oracle, with median difference 0.000~dB. It selects the exact oracle snapshot in 86 of 141 cells, and the largest observed gap is 0.41~dB. These results indicate that validation on held-out received packets is an accurate proxy for selecting snapshots without using the unavailable clean lost packets.

\subsubsection{Causal algorithm deployment}
\label{sec:parcnet-causal}

The causal best-val mode uses only snapshots that have already been validated at the time each lost packet is concealed. On the out-of-domain speech condition, it trails two-pass best-val by approximately 0.1--0.2~dB in packet NMSE, for example \(-1.63\) versus \(-1.84\)~dB at 10\% packet loss. On the in-domain MAESTRO file, causal best-val remains close to both frozen and two-pass best-val. Thus, most of the single-pass adaptation gain is available even when future validation information is not used to reconstruct past losses.

\begin{figure}[t]
  \centering
  \includegraphics[width=\columnwidth]{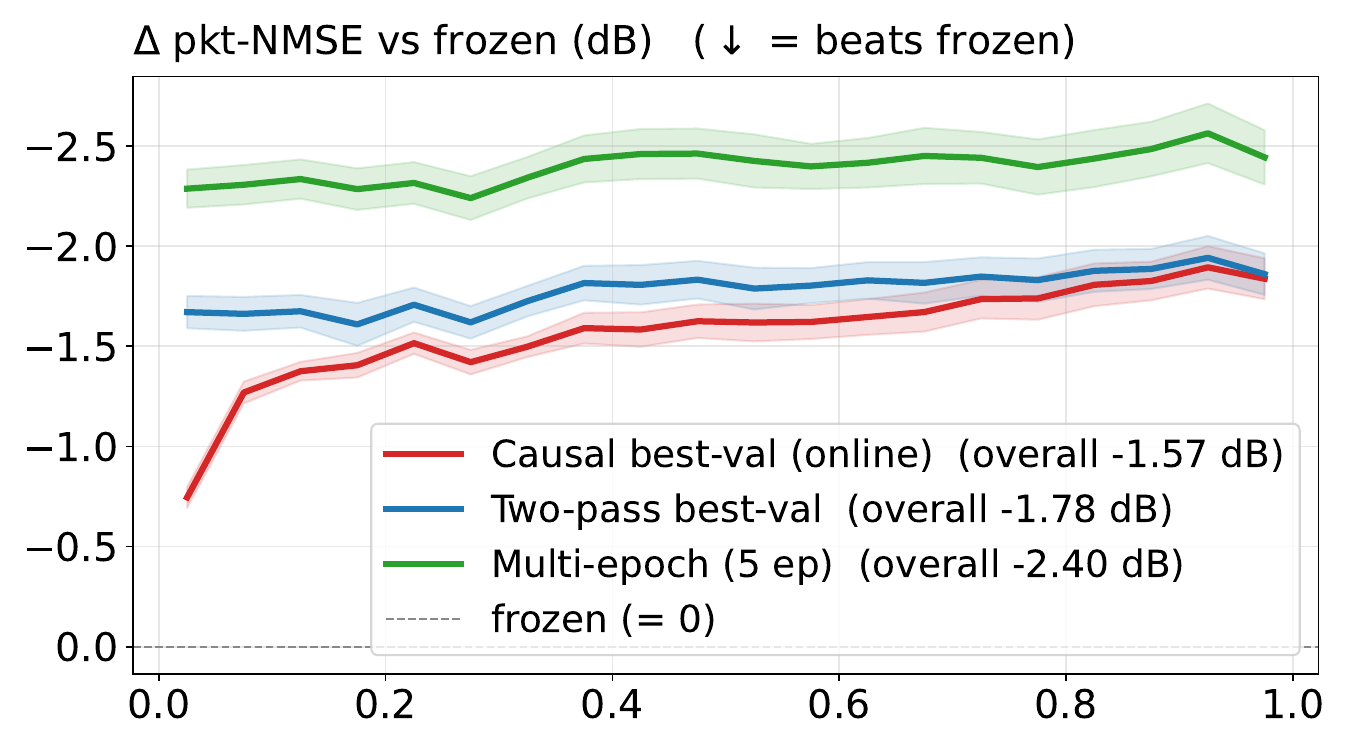}
  \caption{Per-file TTT concealment gain over normalized file position on out-of-domain LibriSpeech at \(r=0.10\), averaged over 40 speakers. Values are packet-NMSE changes relative to frozen PARCnet, so lower is better. Causal best-val has a cold-start cost near the beginning of the file because only early snapshots are available, but it approaches the two-pass result by the end. Multi-epoch adaptation gives the strongest improvement throughout the file. Shaded regions denote \(\pm\) one standard error across files.}
  \label{fig:parcnet_overtime}
\end{figure}

Figure~\ref{fig:parcnet_overtime} shows the position-dependent behavior on out-of-domain LibriSpeech at \(r=0.10\). The causal curve starts closer to frozen because only early snapshots have been validated, but it improves as more received packets become available and approaches the two-pass curve near the end of the file. The two-pass and multi-epoch curves are comparatively stable over file position because they use selected checkpoints for reconstruction over the full file, with multi-epoch providing the largest packet-NMSE improvement throughout.

\subsubsection{Multi-epoch adaptation}
\label{sec:parcnet-multiepoch}

Running adaptation for five epochs and selecting the best snapshot across all epochs gives the strongest completed PARCnet configuration when offline adaptation time is available. On the held-out validation packets, the best snapshot across all epochs improves median packet NMSE by \(-0.40\)~dB relative to the single-epoch best snapshot, with a 95\% confidence interval of \([-0.48,-0.35]\)~dB. The selected snapshot lies in epochs 4--5 for approximately 95\% of cells, indicating that the validation trajectory is still improving beyond the first pass.

The validation improvement also transfers to the true lost packets in the completed multi-epoch cells. Relative to the single-epoch best-val checkpoint from the same adaptation run, the across-epoch selected checkpoint improves test packet NMSE by about 0.5~dB on average and improves every setting. Table~\ref{tab:ood} shows this effect on out-of-domain speech, where multi-epoch best-val gives lower packet NMSE than both single-pass best-val modes at all three packet-loss rates. Table~\ref{tab:ind} shows the same pattern on the in-domain MAESTRO file: multi-epoch best-val gives the lowest packet NMSE at all three loss rates, although the gains are smaller than in the out-of-domain speech condition and the perceptual metrics are less consistent. We therefore treat single-pass best-val as the lower-cost deployment option and multi-epoch best-val as the offline quality-oriented option.

\subsubsection{Metric-specific behavior}
\label{sec:parcnet-where}

The clearest PARCnet gains appear in packet NMSE, which is computed directly on the concealed packets and is also the validation criterion used for snapshot selection. Mel-SC changes are smaller because the metric is computed over the signal representation and is influenced by the large fraction of audio that was received rather than concealed. The perceptual metrics are useful complementary diagnostics but are not always monotonic with packet NMSE. For example, multi-epoch adaptation gives the strongest packet-NMSE reduction on out-of-domain speech, while the PEAQ and PLCmos differences between single-pass and multi-epoch settings are smaller. We should note though, that PLCmos was trained on speech, so it might not be very reliable for music and PEAQ was not developed for PLC purposes so it likewise might not be a proper metric \cite{Mezza2024PARCnet}. 

\subsubsection{Runtime}
\label{sec:parcnet-rtf}

On a single GPU for an 8-minute out-of-domain file at \(r=0.20\), frozen PARCnet concealment has real-time factor 0.075, corresponding to approximately \(13\times\) faster-than-real-time processing. The computational cost of the proposed method is dominated by adaptation: one streaming pass with one update per packet has real-time factor 3.10. Since this cost is dominated by gradient updates, reducing the update frequency provides a direct path toward real-time adaptation. For example, updating once every three packets would reduce the expected adaptation cost to approximately \(3.10/3 \approx 1.03\) real-time factor, before any implementation-level optimization.

The two-pass and held-out-selection settings should therefore be viewed as offline or amortized adaptation procedures in the current implementation: the model is adapted once per file, and the selected checkpoint can then be used for concealment. Multi-epoch adaptation multiplies the adaptation cost by the number of epochs; five epochs correspond to approximately real-time factor 15. This tradeoff is appropriate for offline restoration or quality-oriented processing, while lower update rates are a natural direction for real-time deployment.

\subsubsection{No-leak verification}
\label{sec:parcnet-noleak}

The no-leak constraint is enforced in both the data split and the loss computation. The held-out set \(\mathcal{H}\) is sampled only from observed packets and is never used for gradient updates. At the start of each epoch, the output buffer is re-zeroed at the true loss positions \(\mathcal{L}\). When a true lost packet is encountered, the buffer receives the model prediction, never the clean target audio. In addition, the supervised 80-sample look-ahead region is masked out wherever it overlaps a true lost packet, so clean samples from \(\mathcal{L}\) cannot enter either the MSE or MR-STFT loss.

We verify this implementation with a bit-exact canary test. In the canary run, all clean audio samples at true lost positions are replaced with loud noise before retraining. Under single-threaded deterministic execution, the adapted weights and the complete validation trajectory are byte-identical to the clean-audio run through five epochs. The clean-versus-clean nondeterminism floor and the clean-versus-corrupted-loss signal are both exactly zero. This confirms that clean audio at true lost positions has no path into training, validation, or model selection.

\begin{table}[t]
\centering
\caption{PARCnet results on out-of-distribution LibriSpeech-40. Entries are means over 40 speaker files per loss rate. pkt-NMSE and Mel-SC are lower-better; PEAQ ODG and PLCMOS are higher-better.}
\label{tab:ood}
\scriptsize
\setlength{\tabcolsep}{3.0pt}
\resizebox{\columnwidth}{!}{%
\begin{tabular}{ll rrrr}
\toprule
PLR & Method & pkt-NMSE$\downarrow$ & Mel-SC$\downarrow$ & PEAQ$\uparrow$ & PLCMOS$\uparrow$ \\
\midrule
\multirow{6}{*}{0.10}
  & Zero-fill & $+0.00$ & $0.2424$ & $-3.91$ & $2.43$ \\
  & Frozen & $-0.06$ & $0.1911$ & $-3.64$ & $2.78$ \\
  & \textbf{Causal best-val (ours)} & $-1.63$ & $0.1928$ & $-2.96$ & $2.86$ \\
  & \textbf{Two-pass best-val (ours)} & $-1.84$ & $0.1869$ & $-2.98$ & $2.95$ \\
  & \quad Oracle (single-pass UB) & $-1.85$ & $0.1865$ & $-2.99$ & $2.95$ \\
  & \textbf{Multi-epoch (ours)} & $-2.46$ & $0.1671$ & $-3.28$ & $3.21$ \\
\midrule
\multirow{6}{*}{0.20}
  & Zero-fill & $+0.00$ & $0.3421$ & $-3.91$ & $1.36$ \\
  & Frozen & $+0.10$ & $0.2721$ & $-3.88$ & $1.47$ \\
  & \textbf{Causal best-val (ours)} & $-1.57$ & $0.2739$ & $-3.75$ & $1.47$ \\
  & \textbf{Two-pass best-val (ours)} & $-1.76$ & $0.2673$ & $-3.75$ & $1.51$ \\
  & \quad Oracle (single-pass UB) & $-1.77$ & $0.2670$ & $-3.75$ & $1.51$ \\
  & \textbf{Multi-epoch (ours)} & $-2.34$ & $0.2396$ & $-3.81$ & $1.76$ \\
\midrule
\multirow{6}{*}{0.30}
  & Zero-fill & $+0.00$ & $0.4412$ & $-3.91$ & $1.82$ \\
  & Frozen & $+0.30$ & $0.3569$ & $-3.90$ & $2.09$ \\
  & \textbf{Causal best-val (ours)} & $-1.39$ & $0.3610$ & $-3.83$ & $2.08$ \\
  & \textbf{Two-pass best-val (ours)} & $-1.59$ & $0.3534$ & $-3.83$ & $2.11$ \\
  & \quad Oracle (single-pass UB) & $-1.62$ & $0.3536$ & $-3.83$ & $2.10$ \\
  & \textbf{Multi-epoch (ours)} & $-2.04$ & $0.3270$ & $-3.85$ & $2.27$ \\
\bottomrule
\end{tabular}%
}
\end{table}

\begin{table}[t]
\centering
\caption{PARCnet results on in-distribution MAESTRO solo piano, 50-minute file (single file per loss rate). pkt-NMSE and Mel-SC are lower-better; PEAQ ODG and PLCMOS are higher-better.}
\label{tab:ind}
\scriptsize
\setlength{\tabcolsep}{3.0pt}
\resizebox{\columnwidth}{!}{%
\begin{tabular}{ll rrrr}
\toprule
PLR & Method & pkt-NMSE$\downarrow$ & Mel-SC$\downarrow$ & PEAQ$\uparrow$ & PLCMOS$\uparrow$ \\
\midrule
\multirow{6}{*}{0.10}
  & Zero-fill & $+0.00$ & $0.2410$ & $-3.88$ & $1.18$ \\
  & Frozen & $-5.28$ & $0.1356$ & $-2.32$ & $1.32$ \\
  & \textbf{Causal best-val (ours)} & $-5.52$ & $0.1348$ & $-2.44$ & $1.33$ \\
  & \textbf{Two-pass best-val (ours)} & $-5.59$ & $0.1328$ & $-2.39$ & $1.32$ \\
  & \quad Oracle (single-pass UB) & $-5.61$ & $0.1324$ & $-2.45$ & $1.33$ \\
  & \textbf{Multi-epoch (ours)} & $-5.94$ & $0.1268$ & $-2.36$ & $1.34$ \\
\midrule
\multirow{6}{*}{0.20}
  & Zero-fill & $+0.00$ & $0.3404$ & $-3.78$ & $1.17$ \\
  & Frozen & $-4.92$ & $0.1974$ & $-3.51$ & $1.29$ \\
  & \textbf{Causal best-val (ours)} & $-5.12$ & $0.1961$ & $-3.55$ & $1.30$ \\
  & \textbf{Two-pass best-val (ours)} & $-5.08$ & $0.1956$ & $-3.54$ & $1.29$ \\
  & \quad Oracle (single-pass UB) & $-5.23$ & $0.1965$ & $-3.56$ & $1.30$ \\
  & \textbf{Multi-epoch (ours)} & $-5.49$ & $0.1867$ & $-3.55$ & $1.32$ \\
\midrule
\multirow{6}{*}{0.30}
  & Zero-fill & $+0.00$ & $0.4403$ & $-3.88$ & $1.71$ \\
  & Frozen & $-4.28$ & $0.2679$ & $-3.66$ & $1.43$ \\
  & \textbf{Causal best-val (ours)} & $-4.40$ & $0.2700$ & $-3.67$ & $1.44$ \\
  & \textbf{Two-pass best-val (ours)} & $-4.43$ & $0.2689$ & $-3.67$ & $1.43$ \\
  & \quad Oracle (single-pass UB) & $-4.53$ & $0.2721$ & $-3.65$ & $1.42$ \\
  & \textbf{Multi-epoch (ours)} & $-4.55$ & $0.2700$ & $-3.66$ & $1.42$ \\
\bottomrule
\end{tabular}%
}
\end{table}
\section{Conclusions}

We presented \method, a self-supervised test-time tuning framework for packet loss concealment that adapts pretrained PLC models using only the packets received in the current lossy signal. The method requires no clean reference signal, no external adaptation data, and no architectural changes to the underlying PLC model.

Across two open-source PLC backbones, one for speech and one for music, we showed that received-packet self-supervision can improve concealment on the same test file. The framework is effective in both non-causal and causal deployment settings, and it remains useful across in-domain and out-of-domain signals. These results show that test-time tuning can make existing PLC models more responsive to the current speaker, instrument, recording, and loss pattern.

More broadly, we show that PLC systems do not need to remain fixed at inference time. The packets that arrive at the receiver contain signal-specific information that can be used immediately to adapt the concealer. This provides a practical path toward file-adaptive PLC without retraining on new datasets or modifying model architectures.

% ---------- References ----------
\bibliographystyle{IEEEtran}
\bibliography{refs_canonical_pass2}

\end{document}